\documentclass[preprint,showpacs,amsmath,amssymb]{revtex4}
\usepackage{graphicx}
\usepackage{color} 

\begin{document}

\title{Crumpling wires in two dimensions}
\author{Y. C. Lin, Y. W. Lin, and T. M. Hong}
\affiliation{Department of Physics, National Tsing Hua University, Hsinchu 30043, Taiwan, Republic of China}
\date{\today}

\begin{abstract}
An energy-minimal simulation is proposed to study the patterns and mechanical properties of elastically crumpled wires in two dimensions. We varied the bending rigidity and stretching modulus to measure the energy allocation, size-mass exponent, and the stiffness exponent. The mass exponent is shown to be universal at value $D_{M}=1.33$. We also found that the stiffness exponent $\alpha =-0.25$ is universal, but varies with the plasticity parameters $s$ and $\theta_{p}$. These numerical findings agree excellently with the experimental results. 

\end{abstract}

\pacs{62.20.-x, 05.90.+m, 89.75.Da}

\maketitle

\section{Introduction}
The process of crumpling is everywhere in nature and human activities, including the formation of mountains and valleys in tectonics\cite{tectonics}, packing of DNA strands in viruses\cite{dna}, car wreckage after an accident\cite{car},  or the noisy food wraps that drive us nut in the theater, etc.. In spite of its ubiquity, the mechanism behind many of its properties has remained unclear\cite{witten,witten2}. For instance, how does the labyrinthian internal structure evolve such that it can withstand extraordinary pressure while more than 80\% of its interior remains vacant? Also, why is it that there exists a power law between the external force and the sphere radius with an exponent that varies with material\cite{lin} but, otherwise, is insensitive to the thickness and size of the thin sheet?

\begin{figure}
\includegraphics[width=0.2\textwidth]{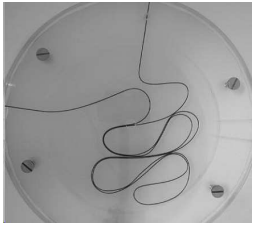}
\caption{ A crumpled copper wire in a two-dimentional cavity. This photo is reproduced by courtesy of Donato {\it et al.}\cite{gomes}.}
\label{fig:photo}
\end{figure}

Scientists have studied crumpled wires (CW) both theoretically\cite{morph,cw1,cw2} and experimentally\cite{gomes,gomes2,gomes3}. As is shown in Fig.\ref{fig:photo}, the wire is smooth and rid of the complicated ridges and vertices, which is different from a crumpled sheet\cite{network}. In the last few years, more and more interesting properties of CW have emerged. For instance, Donato {\it et al.} found a scaling law in the size-mass relation\cite{gomes}, while Stoop {\it et al.} offered the associated morphological phase diagram and reported a power law for  the number of loops with different exponents in each morphology\cite{morph}. However, there are still few quantitative studies of CW either on the energy aspect or macroscopic properties. Furthermore, the former authors all worked on stuffing the wire into a fixed two-dimensional cavity. It is not clear whether this shares the same properties as by increasing the strength of the confining potential while fixing the wire length. Aside from this, another motivation for us is to check how sensitive is the CW on the specific form and relative amplitude of stretching and bending energies. In this report, simulations are carried out to study the mechanical properties of CW under the semiflexible polymer model\cite{cw2} and the minimal energy model\cite{ens}. To prevent the entropy from dominating the statistical behavior\cite{degennes}, the wire is chosen to be of medium length.

\section{Model and computational method}
In our simulations, the optimal shape of a wire is obtained by minimizing the total energy of CW. We begin with a wire in a weak external field, and gradually increase the field strength. At each stage with a certain field strength, we recursively minimize the energy and move to the next stage with a larger external field after an equilibrium state has been achieved. This process is carried out throughout the whole simulation until the wire structure no longer changes with the increasing external field. The wire consists of {\it N} number of monomers and is of length $L_0=1$. The total energy of a CW includes bending energy $E_{b}$, stretching energy $E_{s}$, and external potential $E_{ext}$. In addition, a hard-core potential $E_{hc}$ is incorporated to keep the wire from self-crossing. Explicitly, the energy function can be written as:
\begin{equation}
E_{tot}=\sum_{i=1}^{N-1}\frac{k_{b}}{2}\theta_{i}^{2}+\sum_{i=1}^{N}\Big[\frac{k_{s}}{2}(\Delta l_{i})^{2}+\lambda r_{i}^{2}\Big]+E_{hc}
\label{eq:model}
\end{equation}
where $\theta_{i}$ is the bending angle between two successive segments and $k_{b}=YI/(L_{0}/N)$ is the bending rigidity which is proportional to the Young's modulus $Y$ and second moment of inertia $I$. The stretching modulus $k_{s}=YA/(L_{0}/N)$ depends on the wire cross section $A$, and $\Delta l_{i}$ is the deviation of the $i$-th segment from its equilibrium length $L_{0}/N$. The distance $r_i$ of the {\it i}-th monomer is measured from the the center of the external parabolic potential. Since most research interests fall into cases with insignificant extension, we mainly focused on conditions that $k_{b}\ll k_{s}$. Typical parameters for simulations are listed in Tab.~\ref{tb:parameter}. The steady state of CW is determined by minimizing its total energy with Powell's algorithm. To make sure that the local minimum we found was not a special case, we adopted a random set of searching directions in the Powell's algorithm and repeated the simulations for ten times to get the averaged results. Most of the time, CW are randomly folded rather than forming a perfect spiral - the apparent global minimum\cite{spiral}. This indicates that those folded configurations are metastable states. The modified Powell's minimization is equivalent to the Monte Carlo simulation in this aspect.

\begin{table}[t]
\caption{The value of parameters used in simulations}
\label{tb:parameter}
\begin{ruledtabular}
\begin{tabular}[t]{cc}
Parameter& Range of values   \\\hline
Bending rigidity $k_{b}$ & $1$ to $10$\\
Stretching modulus $k_{s}$&$10^{4}$ to $10^{7}$\\
External field $\lambda$&1 to $2 \times10^{3}$\\
Number of segments $N$&100\\
Wire length $L_{0}$& 1\\
Width of hard core& $10^{-5}$\\
\end{tabular}
\end{ruledtabular}
\end{table}
 
\begin{figure*}
\includegraphics[width=0.8\textwidth]{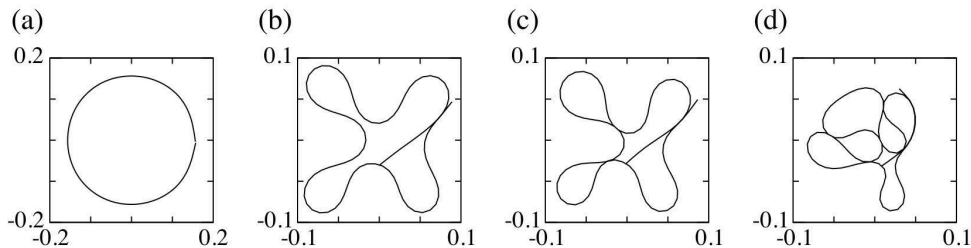}
\caption{A sequence of simulation results for $k_{b}=5$ and $k_{s}=5 \times 10^{5}$. The value of $\lambda$ for (a), (b), (c), and (d) are $70,200,300,$ and $1000$ respectively. It can be seen that when we increase $\lambda$, no sharp edge occurs and the shape of CW remains smooth. Those local structures start to contact each other when $\lambda \geq 1000$. More turning points are formed if the external field increases.}
\label{fig:demon}
\end{figure*}

\section{Results and discussion}
\subsection{shape and size-mass exponent}
The loops of CW in Figs.\ref{fig:photo} and \ref{fig:demon} resemble the boundary of a water-drop-like structure. Compare with the simulations for a semi-flexible polymer\cite{cw2} which explicitly put in finite temperatures and consequently resulting in many small fluctuations, the configuration of CW is characteristically smooth, and so we exclude the temperature effect in this simulation. We believe the smoothness is credited to the stretching force which causes the relaxation of roughness, analogous to the surface tension which disfavors a kinky surface. Because the stretching energy is found to be insignificant when $k_{b} \ll k_{s}$, we can neglect it when determining the mathematical form of the configuration. For simplicity and without the loss of generality, we consider the case of $N\rightarrow \infty$, i.e., the continuous limit. Eq.(\ref{eq:model}) then can be re-written as:
\begin{eqnarray}
E_{tot}^{*}= \int^L_0 \Big[\frac{k_{b}}{2}\big(\frac{d \theta}{ds}\big)^2+ \lambda r^{2}\Big] ds
\label{eq:lagrangian}
\end{eqnarray}
The search for shape function ${r}(\theta )$ with minimal energy is done by introducing the variational method. Fix both $ {r}(s=0)$ and $ {r}(s=L)$ and this total energy is identical to the classical action once we identify the arc length parameter $s$ as the time. The Euler-Lagrange equation of motion immediately gives out the shape function as $r^{2} \sqrt{k_{b}/ \lambda}=\sin 2 \theta$, where the $r^2$ term is not hard to expect by the dimensional analysis in retrospect. This matches the pattern in the initial stage of the crumpling observed experimentally\cite{gomes} as well as in our simulations. However, when the loops start to touch each other as the crumpling proceeds further, the excluded volume or steric interaction\cite{degennes} is expected to squeeze and distort the shape of loops.

\begin{figure}
\includegraphics[width=0.35\textwidth]{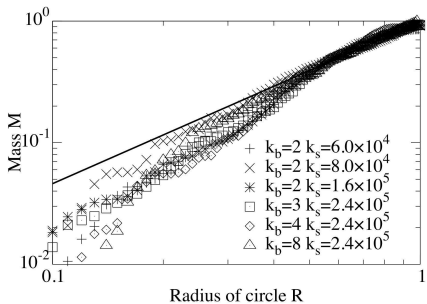}
\caption{ The mass-size relation of CW for $\lambda$=1000. The radius of the circle is rescaled by $R_{G}$. Apparently, all mass will already be included in the circle when $R=1$. The rescaling process only shifts the data on a Log-Log plot and does not affect the value of exponent $D_{M}=1.33$. Since we set the origin of the circle at the center of mass of the CW, there is a bias in the data at small R. This bias is avoided by fitting only from $R$=0.5 to $R$=1 as shown by the solid straight line\cite{gomes}.}
\label{fig:rmdemon}
\end{figure}

The mass function $M(R)$ is a common property\cite{gomes} to characterize the configuration of CW. It is defined as the total mass encompassed within a circular area of radius $R$ centering at the CW's center of mass:
\begin{eqnarray}
M(R)=\sideset{}{}\sum_{i=1}^{N-1}\theta(R-r_{i})\delta(\theta(r_{i}-R),\theta(r_{i+1}-R))\frac{1}{N}\\ \nonumber
+\delta(\theta(r_{i}-R),\theta(R-r_{i+1}))\frac{1}{N}\: f(r_{i},r_{i+1})
\label{eq:sm}
\end{eqnarray}
where $\theta (x)$ denotes the step function, $\delta (x,y)$ stands for the Kronecker delta-function which equals 1 when $x=y$ and zero otherwise, and $f(r_{i},r_{i+1})$ is the portion of length between $r_{i}$ and $r_{i+1}$ within the circle.
After performing a series of simulations, the mass-size relation is computed and shown in Fig.\ref{fig:rmdemon} for six distinct combinations of $k_b$ and $k_s$. We can see that the mass of CW grows in a power-law fashion when the radius increases and, disregarding\cite{gomes} the fluctuations at small external field, the growing exponent saturates at $D_M=1.33$ for $\lambda >100$, see Fig.\ref{fig:rmdemon}. The saturating behavior is in accordance with Donato {\it et al.}'s conjecture\cite{gomes} that in a loose-packing situation the corresponding mass-size exponent has insignificant dependence on how the wires are injected. And the value of $D_{M}=1.33$ is also consistent with Donato {\it et al.}'s result.

\subsection{Energy allocation}
\begin{figure}
\includegraphics[width=0.45\textwidth]{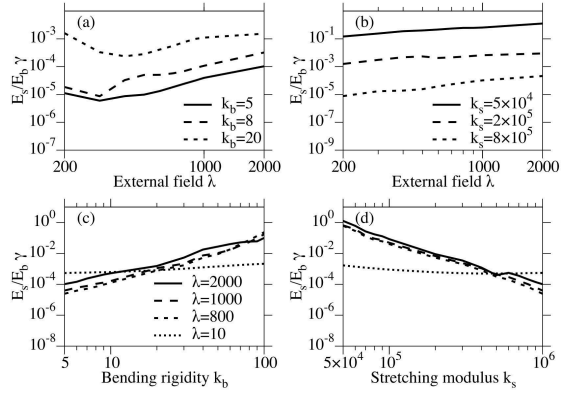}
\caption{The value of $\gamma$ is plotted against several parameters. For (a) and (b), $\gamma$ grows with the strength of external field in a power-law fashion with exponent $\epsilon_{l}=0.55$. In (a) the stretching modulus is set at $2\times 10^{3}$, and in (b) the bending rigidity is $6$.  Panels (c) and (d) depict $\gamma$ versus $k_{b}$ and $k_{s}$ respectively, and they share the same denotation of data. Notice that the power-law can only be found for strong external field ($\lambda \gtrsim 500$). The exponents of the power-law are determined to be $\epsilon_{b}=2.6$ and $\epsilon_{s}=-3.3$}
\label{fig:energy}
\end{figure}

We characterize the allocation of elastic energies in CW by the ratio $\gamma$=$E_{s}/E_{b}$. Data in Fig.~\ref{fig:energy} establish a scaling relation between $\gamma$ and the three parameters in the Hamiltonian as:
\begin{eqnarray}
\gamma \sim \lambda^{\epsilon_{l}}\medspace k_{b}^{\epsilon_{b}}\medspace k_{s}^{\epsilon_{s}}
\label{eq:scaling}
\end{eqnarray}
where $\epsilon_{l}=0.55$, $\epsilon_{b}=2.6$, and $\epsilon_{s}=-3.3$. The numerical results can be deduced analytically. These exponents can be obtained by doing a dimensional analysis on the bending energy and the external potential in Eq.(\ref{eq:lagrangian}), $k_b/\ell\sim \lambda \ell^3$, which gives the characteristic size for each loop as $\ell\sim(k_b/\lambda )^{1/4}$. Multiplying the bending energy stored in each loop, $k_b/\ell$, by the number of loops $L/\ell$ enable us to estimate the bending energy as $k_b L/\ell^2$. In a similar fashion, the stretching energy can be written as $E_s\sim k_s (L-L_0)^2$. Minimizing the sum of these two energies with respect to $L$ gives
\begin{equation}
\gamma =\frac{\sqrt{k_b\lambda}}{k_s\big[L_0-\sqrt{k_b \lambda}/k_s\big]}
\end{equation}
 The value of $\epsilon_{l}=0.5$ matches the simulation result. Although the exact magnitudes are not the same, the signs of $\epsilon_{b}$ and $\epsilon_{s}$ and their relative magnitude are still captured. One possible explanation for the discrepancy is our failure to include the hard-core potential in the analytic arguments, which has been known to be difficult\cite{degennes} but crucial for a realistic polymer.

\begin{figure}
\includegraphics[width=0.35\textwidth]{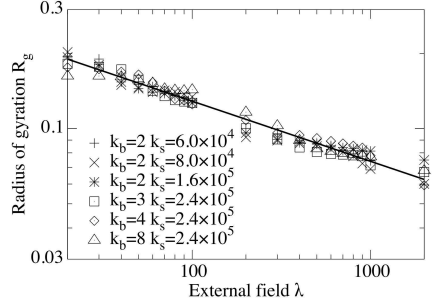}
\caption{The shrinkage of $R_{G}$ versus the strength of confining potential for CW with different rigidity and modulus, $k_{b}$ and $k_{s}$. All data collapse to the parameters $k_{b}=8,k_{s}=2.4\times 10^{5}$ with  the scaling $R_{g}\sim (k_{b} k_{s})^{1.2}$. The solid line is the fitting result $R_{g}\sim \lambda^{-0.25}$.}
\label{fig:size}
\end{figure}

\subsection{Stiffness and plasticity}
How to estimate the resistance of a crumpled structure is a well-known difficult problem. For instance, what is the main reason why such a loose crumpled structure can be so hard? As a starting point, people has studied the power law between the structure size and the applied force both experimentally and theoretically\cite{nagel,lin}. Our simulation result in Fig.\ref{fig:size} showed that the radius of gyration $R_{g}=r_{max}$ of CW decreases with $\lambda$ in a power-law form. To prevent the structure from becoming too complicated, we limit the field strength $\lambda$ at 2000 to optimize the resolution and accuracy. The data are found to collape into the scaling relation $R_{g}\sim \lambda^{\alpha}(k_{b} k_{s})^{1.2}$, which suggests that the stretching energy not only correlates those monomers, but also contributes to the stiffness. Intuitively, the dependence of $k_s$ is expected to drop out instead of diverging as it approaches infinity. The reason why we obtained such a scaling relation is a consequence of the discrete model since the length of monomers is bound to be compressed as they all scramble to be near the center of the external potential. 

In Fig.\ref{fig:size}, $\alpha=-0.25$ is determined to be independent of $k_{b}$ and $k_{s}$ and, consequently, the thickness and Young's modulus of the wire. This result is the same as the predictions made for an elastic thin sheet\cite{nature}. It can be derived analytically by minimizing with respect to $R$ the sum of the external potential, $\lambda R^2L_0$, and the total repulsive interaction, $\beta (L_0/R^2)^2\ R^2$ where $\beta$ is a measure of the hard-core potential, $L_0/R^2$ is the monomer density, and the other $R^2$ factor comes from integrating over the cavity.

\begin{figure}
\includegraphics[width=0.35\textwidth]{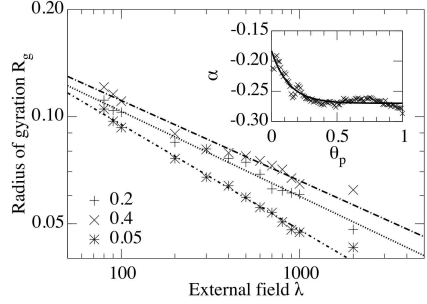}
\caption{ The $\lambda$-$R_{g}$ relation of the CWs($(k_{b},k_{s})=(5,5\times 10^{5})$) with plasticity for different $s$ which are denoted in the figure. The value of the exponents $\alpha$ are -0.21, -0.24, and -0.29 for s=0.4, 0.2, and 0.05 respectively. The inset illustrates that $\alpha$ decays with $\theta_{p}$ exponentially and eventually saturates at value of $0.26$ for $s=0.4$.}
\label{fig:plasticity}
\end{figure}

Since Stoop {\it et al.} has demonstrated the importance of plasticity for CW, we also include it in our simulation. The bending rigidity $k_{b}$ is revised with a linear approximation, $sk_{b} (0 \leq s \leq 1)$ in the plastic regime, so that the bending energy becomes $E_{b}= k_{b} \theta _{p}^{2}/2 +sk_{b}\big(\theta^2 - \theta _{p}^2\big)/2+k_{b} \theta _{p}(1-s)(\theta -\theta _{p})$ when $\theta$ is larger than the yield threshold angle $\theta_{p}$\cite{morph}. We analyzed the $\lambda$-$R_{g}$ relation of CW for $s=0.05, 0.1, 0.2, 0.4,$ and $0.8$, and it behaves very different from that of the elastic one. In Fig.\ref{fig:plasticity}, the value of $\alpha$ decays with the bending rigidity in the plastic regime which is consistent to the experimental findings in \cite{lin}. It also decays exponentially with the yield threshold $\theta_{p}$. When large deformations cost less energy due to the plasticity, they appear in abundance and cause many vertex-like structures which makes the CW stiffer.

\section{Conclusion}
In conclusion, we introduced a minimal one-dimensional model to simulate the crumpled wires. The simulation results agree excellently with the experimental observation, which include the pattern of loops, the universal value of mass-size exponent $D_{M}=1.33$\cite{gomes}. We checked the exponent $\alpha$ that characterizes the power-law relation between the crumping force and the CW radius to be independent of the bending rigidity and the stretching modulus which is consist to \cite{nature}. Plasticity is found to suppress $\alpha$ which is in the right trend as the recent experiment\cite{lin} on crumpled sheets.

We benifit from correspondence and discussions with Professors P. C. Chen,  H. J. Herrmann, P. Y. Hsiao, Y. Kantor, H. H. Lin, and E. Terentjev. Support by the National Science Council in Taiwan under grant 95-2120-M007-008 is acknowledged.



\end{document}